\documentclass[eqsecnum,aps,twocolumn,epsf]{revtex4} 
\usepackage{graphicx}
 
\begin{document}
   
 

\title{Mean-field description of dynamical collapse of a fermionic
condensate in a trapped 
boson-fermion mixture}
 
\author{Sadhan K. Adhikari}
\affiliation
{Instituto de F\'{\i}sica Te\'orica, Universidade Estadual
Paulista, 01.405-900 S\~ao Paulo, S\~ao Paulo, Brazil\\}

\date{\today}
 
 
\begin{abstract}

We suggest a time-dependent dynamical mean-field-hydrodynamic model for
the collapse of a trapped boson-fermion condensate and perform numerical
simulation based on it to understand some aspects of the experiment by
Modugno {\it et al.} [Science {\bf 297}, 2240 (2002)] on the collapse of
the fermionic condensate in the $^{40}$K-$^{87}$Rb mixture. We show that
the mean-field model explains the formation of a stationary boson-fermion
condensate at zero temparature with relative sizes compatible with
experiment. This model is also found to yield a faithful representation of
the collapse dynamics in qualitative agreement with experiment.  In
particular we consider the collapse of the fermionic condensate associated
with (a) an increase of the number of bosonic atoms as in the experiment
and (b)  an increase of the attractive boson-fermion interaction using a
Feshbach resonance. Suggestion for experiments of fermionic collapse using
a Feshbach resonance is made.

\end{abstract}

\maketitle

\section{Introduction}
 
Recent successful observation   of condensed boson-fermion
mixtures  of trapped alkali atoms by different experimental groups
\cite{exp1,exp2,exp3,exp4} 
has initiated the  intensive experimental studies of different novel
phenomena \cite{exp5,exp5x,exp6}. Among these experiments there have been 
studies of condensate of two components of $^{40}$K \cite{exp1} and  
$^6$Li \cite{exp2} atoms. Condensation of boson-fermion  mixtures 
$^{6,7}$Li \cite{exp3}, $^{23}$Na-$^6$Li \cite{exp4} and 
$^{87}$Rb-$^{40}$K \cite{exp5,exp5x} have also been reported. The
collapse of fermionic condensate in a boson-fermion mixture
$^{87}$Rb-$^{40}$K 
has been
observed and studied by Modugno {\it et al.} \cite{exp5}. In this paper we
consider the collapse dynamics of fermions in a boson-fermion
mixture using a coupled time-dependent mean-field-hydrodynamic model. The
bosonic
part is
treated by the mean-field Gross-Pitaevskii (GP) equation \cite{11} and the
fermionic
part is treated by a hydrodynamic model. The present 
mean-field-hydrodynamic model could be considered to be a time-dependent
extension of a time-independent model
suggested for the stationary equilibrium states by Capuzzi {\it et al.}
\cite{capu}.

Because of the Pauli exclusion principle the fermions in spin parallel
states in the condensate experience a strong repulsion. This is the
dominating interaction at short distances and avoids the collapse of a
fermionic condensate. In contrast an 
attractive bosonic condensate larger than a critical size is not
dynamically
stable  \cite{hulet}. However, if such a bosonic 
condensate is ``prepared"  or somehow made to exist it experiences a
dramatic collapse and explodes
emitting atoms leading to a condensate
of small size with a high
central density. Under high pressure three-body recombination takes place 
with the emission of energy leading to explosion and loss of atoms.
The possibility of
collapse was
first suggested    and observed in $^7$Li atoms \cite{hulet}.
A dynamical study of controlled collapse   and explosion
has been performed by Donley {\it et al.}
\cite{don} on an attractive  $^{85}$Rb bosonic condensate, where they
manipulated the inter-atomic interaction by changing the external magnetic
field exploiting  a nearby Feshbach resonance \cite{fs}. 
In the vicinity
of a Feshbach resonance the atomic scattering length $a$ can be varied
over a huge range by adjusting an external magnetic field, thus
transforming a repulsive  bosonic condensate to a highly attractive one.  
There
have been many  theoretical \cite{th1,th2} studies to describe different
features of the experiment by Donley {\it et al.} \cite{don}.

In the classic study of a trapped boson-fermion mixture   Modugno
{\it et
al.} \cite{exp5}  simulated a situation of a strong  effective interaction
between fermions via a strongly attractive boson-fermion interaction which
may play an important role in the collapse of the fermionic condensate. 
The presence of a strong attractive interaction between bosons and
fermions can induce an effective attraction between the fermions
\cite{xxx}. If this overall effective attraction among the fermions can
overcome the Fermi repulsive pressure there is a possibility of the
collapse of a fermionic condensate in a boson-fermion mixture.  Modugno
{\it et
al.} \cite{exp5} showed in an experiment with  $^{87}$Rb-$^{40}$K mixture
that it is indeed the case and produced some results of the collapsing
dynamics of this system which we would like to understand using a
mean-field-hydrodynamic model. 

Previously, in addition to the study of the collapse of a purely bosonic
condensate \cite{th1,th2},
we also investigated \cite{adhi,adhix,k} the collapse dynamics  in a
two-species
bosonic
condensate
initiated by an attraction between interspecies bosons
using the  mean-field coupled GP equation \cite{11}. It
was
found that
even when the intra-species interaction is repulsive one can have a
collapse in one or two species due to the  interspecies attraction which
simulates an effective  intra-species   attraction. The situation is
qualitatively similar in the present boson-fermion mixture. The
boson-fermion attraction in a boson-fermion mixture can induce  an
attraction in the
fermionic system leading to collapse.
However, there
is no coupled GP equation in the case of boson-fermion mixture which
complicates a theoretical description. Instead, in this paper we use a
time-dependent version of a 
recently suggested equations of generalized hydrodynamics
\cite{capu}. These
equations allow us to span the entire range of boson-fermion interaction
as in the coupled GP equation for bosons \cite{adhi}.

The equilibrium properties and the phase diagram of a boson-fermion
mixture have been studied by several authors \cite{yyy} to obtain good
agreement with experiment using a mean-field-type description \cite{zzz}.
All of them employed a time-independent formulation. To the best of our
knowledge the study of nonequilibrium properties of fermionic collapse
using a time-dependent formulation is considered for the first time in 
this paper. Our findings are in agreement with the experiment by Modugno
{\it et al.} \cite{exp5}.

In Sec. II we present our  time-dependent mean-field model. 
This consists of a set of coupled partial differential equations 
involving the bosonic and fermionic condensate wave functions. 
We introduce an appropriate loss term due to three-body boson-fermion 
recombination. 
In Sec.  III we present
our results for stationary boson-fermion wave functions as well as 
fermion collapse initiated by a growth in boson number and boson-fermion
interaction. Our stationary results are consistent with 
experiment \cite{exp5} and other numerical studies
\cite{capu,yyy,zzz}. The present study also yields a faithful
representation of the collapse dynamics as observed 
experimentally \cite{exp5}.
Finally, 
 a summary of our findings  is given in Sec. IV.
 
\section{Nonlinear mean-field-hydrodynamic model}

The time-dependent Bose-Einstein condensate wave
function $\Psi({\bf r};t)$ at position ${\bf r}$ and time $t $
allowing
for atomic loss
may
be described by the following  mean-field nonlinear GP equation
\cite{11}
\begin{eqnarray}\label{a} \biggr[& -& i\hbar\frac{\partial
}{\partial t}
-\frac{\hbar^2\nabla^2   }{2m_{{B}}}
+ V_{{B}}({\bf r})
+ g_{{BB}} n_B
 \biggr]\Psi_{{B}}({\bf r};t)=0, \nonumber \\
\end{eqnarray}
with normalization $ \int d{\bf r} |\Psi_B({\bf r};t)|^2 = N_B. $ 
Here $m_{{B}}$
is
the mass and  $N_{{B}}$ the number of bosonic atoms in the
condensate, $n_B\equiv  |\Psi_{{B}}({\bf r};t)|^2$ is the boson 
probability density,
 $g_{{BB}}=4\pi \hbar^2 a_{{BB}}/m_{{B}} $ the strength of
inter-atomic interaction, with
$a_{{BB}}$ the  scattering length. 
The trap potential with spherical symmetry may be written as  $
V_{{B}}({\bf
r}) =\frac{1}{2}m_B \omega_B ^2r^2$ where
 $\omega_B$ is the angular frequency.
The probability density of an isolated 
fermionic condensate  $ n_F\equiv |\Psi_F({\bf r})|^2$ is given by
\begin{eqnarray}\label{b}
n_F= \frac{[\epsilon_F-V_F({\bf r})]^{3/2}}{A^{3/2}}
\end{eqnarray}
where $\Psi_F({\bf r})$ is the condensate wave function, $A=\hbar^2 (6
\pi^2
)^{2/3}/ (2m_F)$, $\epsilon_F$ is the Fermi energy, and $m_F$ is the
fermionic mass. The spherical trap is
given by
$V_F({\bf
r}) =\frac{1}{2}m_F \omega_F ^2r^2$. The number of fermionic atoms $N_F$
is given by the normalization $\int d{\bf r} |\Psi_F({\bf r})|^2=N_F$.

However, it has been suggested  that if there is an interaction between 
the fermions and bosons, Eqs. (\ref{a}) and (\ref{b}) get modified to
\cite{zzz} 
\begin{eqnarray}\label{c} \biggr[& -& i\hbar\frac{\partial
}{\partial t}
-\frac{\hbar^2\nabla^2}{2m_{{B}}}
+ V_{{B}}({\bf r})
+ g_{{BB}} n_B \nonumber \\
&+& g_{{BF}} n_F
 \biggr]\Psi_{{B}}({\bf r};t)=0. 
\end{eqnarray}
and 
\begin{eqnarray}\label{d}
n_F= \frac{(\epsilon_F-V_F -g_{{BF}}
n_B)^{3/2}}{A^{3/2}},
\end{eqnarray}
where $g_{BF}=2\pi \hbar^2 a_{BF}/m_R$ with the
boson-fermion reduced mass $m_R=m_Bm_F/(m_B+m_F)$ where
$m_F$ is the mass of fermionic atoms. The interaction between fermions in
spin polarized state is highly suppressed due to Pauli exclusion principle
and has been neglected in Eqs. (\ref{c}) and (\ref{d}) and will be
neglected throughout this paper. Also, we shall always assume
boson-fermion attraction ($a_{BF}<0$) and boson-boson repulsion
($a_{BB}>0$).

Capuzzi {\it et al.} \cite{capu} developed a set of time-independent  
equations
for
boson-fermion condensate from a consideration of hydrodynamic motion of
two condensed fluids, one bosonic and another fermionic, in a spherical
trap at zero temperature. The interaction between bosons and between
bosons and fermions are described by contact potentials parametrized by
coupling constants $g_{BB}$ and $g_{BF}$ defined above. They derived the
following mean-field energy functional for the system \cite{capu}
\begin{eqnarray}\label{xx}
[E]&=& \int d {\bf r} \left(\frac{\hbar^2|\nabla
\Psi_B|^2 
}{2m_B}+V_B|\Psi_B|^2+\frac{1}{2}g_{BB} |\Psi_B|^4\right)\nonumber \\
&+& \int d {\bf r} \left(\frac{\hbar^2 |\nabla
\Psi_F|^2 }{6m_F}+
V_F|\Psi_F|^2+\frac{3}{5}  A  |\Psi_F|^{10/3}\right)\nonumber \\
&+& g_{BF} \int d {\bf r}   |\Psi_F|^2|\Psi_B|^2.  
\end{eqnarray}  
The first integral on the right-hand side of Eq. (\ref{xx}) is the
Gross-Pitaevskii energy functional of the bosons and is related to the
(nonlinear) Schr\"odinger equation \cite{11}. However, the second
integral, although bears a resemblance with the first,  
is derived from the hydrodynamic equation of motion of the fermions 
and is not related to a  Schr\"odinger-like equation \cite{capu}. Hence,
the
derivative term
in the second integral has a different mass factor $6m_F$ and not the
conventional  Schr\"odinger mass factor $2m_B$ as in the first integral.
Finally, the last integral corresponds to an interaction between bosons
and fermions.

Equation (\ref{xx}) corresponds to the following Lagrangian 
density 
\begin{eqnarray}\label{yy}
{\cal L}&=& \frac{i}{2}\hbar \sum_{i=B,F}\left(
\Psi_i\frac{\partial \Psi_i^*}{\partial t} - \Psi_i^*
\frac{\partial
\Psi_i}{\partial
t} 
\right) \nonumber \\ 
&+&  \left(\frac{\hbar^2|\nabla
\Psi_B|^2
}{2m_B}+V_B|\Psi_B|^2+\frac{1}{2}g_{BB} |\Psi_B|^4\right)\nonumber \\
&+&  \left(\frac{\hbar^2 |\nabla
\Psi_F|^2 }{6m_F}+
V_F|\Psi_F|^2+\frac{3}{5}  A  |\Psi_F|^{10/3}\right)\nonumber \\
&+& g_{BF}    |\Psi_F|^2|\Psi_B|^2.
\end{eqnarray}
The mean-field dynamical equations for the system are just the 
usual  Euler-Lagrange  equations 
\begin{equation}
\frac{d}{dt}\frac{\partial {\cal L}}{\partial   \frac{\partial
\Psi_i^*}{\partial t}}+
\sum _{k=1}^3 \frac{d}{dx_k}\frac{\partial {\cal L}}{\partial
\frac{\partial \Psi_i^*}{\partial x_k}}= \frac{\partial {\cal L}}{\partial
 \Psi_i^*},
\end{equation}
where $x_k, k=1,2,3$ are the three space components, 
$i=B$ corresponds to the bosonic wave function and $i=F$ corresponds
to the 
fermionic wave function. With the Lagrangian density (\ref{yy})
these equations of motion become: 
\begin{eqnarray}\label{e} \biggr[ &-& i\hbar\frac{\partial
}{\partial t}
-\frac{\hbar^2\nabla^2}{2m_{{B}}}
+ V_{{B}}({\bf r})
+ g_{{BB}}n_B \nonumber \\
&+& g_{{BF}} n_F
 \biggr]\Psi_{{B}}({\bf r};t)=0. 
\end{eqnarray}
\begin{eqnarray}\label{f} \biggr[& -& i\hbar\frac{\partial
}{\partial t}
-\frac{\hbar^2\nabla^2}{6m_{{F}}}
+ V_{{F}}({\bf r}) 
+ A n_F^{2/3} \nonumber \\
&+& g_{{BF}} n_B
 \biggr]\Psi_{{F}}({\bf r};t)=0. 
\end{eqnarray}
The normalization of the wave functions are given by 
$\int d{\bf r} |\Psi_{{B}}({\bf r};t)|^2 =N_B$ and 
$ \int d{\bf r} |\Psi_{{F}}({\bf r};t)|^2  =N_F.$ It is understood that
Eq. (\ref{d}) results as a consequence of Thomas-Fermi-type approximation
to Eq. (\ref{f}), when the kinetic energy term in the latter (equation) is
neglected at low temperatures. 

Equations (\ref{e}) and (\ref{f})  are the desired time-dependent
mean-field-hydrodynamic equations. 
We shall use these equation for the study of the collapse of
the fermionic atoms in a coupled boson-fermion mixture. If $g_{BF}$ were
zero, both
the bosonic and fermionic condensates would have been stable. An
attractive or negative  $g_{BF}$  can make  one or both systems to
become highly attractive and lead to 
collapse. Similar effect was found before in the study of two coupled
bosonic condensates \cite{adhi}. In that study it was found that even if
the
interaction among atoms of the same species is repulsive an attractive
interaction among atoms of two different species can cause one or both
components of
the coupled
system to collapse. Similarly, in a coupled atomic-molecular condensate a
strong attraction between atoms and molecules can cause one or both of
atomic and molecular condensates to collapse \cite{adhix}.

To study the phenomenon of collapse we have to add the proper mechanism
for atom loss in Eqs. (\ref{e}) and (\ref{f}). The attractive interaction
between a bosonic atom $B$ and a fermionic atom $F$ leads to the following
three-body
recombination process to form a molecule $(BF)$ composed of a bosonic and
fermionic atom, which is responsible for the loss of atoms \cite{exp5} 
\begin{equation} \label{g}
B + B + F   \to   (BF) + B. 
\end{equation}
As the fermions are in a spin-polarized state, due to Pauli principle
two-fermion molecules cannot be formed. 
Consequently, femionic atoms could only be lost in the presence of bosons
and the loss rate scales quadratically with bosonic density
and is
independent of fermion number $N_F$ \cite{exp5}. In addition to reaction
(\ref{g}), there is also the possibility of the formation of a two-boson
molecule $(BB)$ via the reaction 
\begin{equation} \label{g2}
B + B + B   \to   (BB) + B.
\end{equation}
However, for a  low-density {\it repulsive} bosonic condensate as in the
experiment on boson-fermion mixture \cite{exp5}, the rate of loss of
bosons due to
reaction
(\ref{g2}) will presumably be  small \cite{sl}. Moreover, reaction
(\ref{g2}) has no influence on the collapse of the fermionic
condensate. As the primary motivation of this paper is to investigate the
collapse of the fermionic condensate, we neglect reaction
(\ref{g2}) in our study. 

   The
recombination reaction
(\ref{g}) adds 
 an extra term in Lagrangian 
density  (\ref{yy}) of the form 
$-i\hbar K_3 |\Psi_F|^2 |\Psi_B|^4/2$ proportional to the product of
fermion density and the square of boson density, where $K_3$ is the loss
rate. In the presence of only the bosonic atoms   the
contribution to the  Lagrangian density for the recombination reaction
(\ref{g2}) is the usual three-body
recombination
term $-i\hbar K_3 |\Psi|^6 /2$ \cite{th2,sl} proportional to the cube of
boson density. After proper symmetrization, 
with the use of the loss term $-i\hbar K_3 |\Psi_F|^2 |\Psi_B|^4/2$
in Eq. (\ref{yy}),  Eqs. (\ref{e}) and
(\ref{f}) become  
\begin{eqnarray}\label{h}& \biggr[& - i\hbar\frac{\partial
}{\partial t}
-\frac{\hbar^2\nabla^2}{2m_{{B}}}
+ V_{{B}}({\bf r})
+ g_{{BB}} n_B +g_{BF}n_F   \nonumber \\          
&-&{i\hbar}K_3
n_B n_F  
 \biggr]\Psi_{{B}}({\bf r};t)=0,
\end{eqnarray}
\begin{eqnarray}\label{i}& \biggr[& - i\hbar\frac{\partial
}{\partial t}
-\frac{\hbar^2\nabla^2}{6m_{{F}}}
+ V_{{F}}({\bf r})
+ A n_F^{2/3}   
+ g_{{BF}} n_B  \nonumber \\     
&-& {i\hbar}K_3
n_B^2
 \biggr]\Psi_{{F}}({\bf r};t)=0.
\end{eqnarray}
 
In the spherically symmetric state of zero angular momentum 
the wave
function has the form $\Psi({\bf
r};t)=\psi(r;t).$ Of the experimental condensation of boson-fermion
mixtures the two pioneering possibilities were the mixture of $^6$Li
(fermion) and 
$^{23}$Na (boson) \cite{exp4}, and the mixture of $^{40}$K (fermion) and
$^{87}$Rb
(boson) \cite{exp5}. The actual experiment on the collapse of fermions
\cite{exp5} was performed in the $^{40}$K-$^{87}$Rb system 
with an axially symmetric trap. In this
paper
we shall  consider the effect of the variation of the boson-fermion
scattering
length $a_{BF}$ on the collapse. For that purpose one requires not only an
accurate value of $K_3$ but a variation of $K_3$ with 
$a_{BF}$. As the dynamics of collapse is sensitive to this unknown
variation of  $K_3$, we are
not in a position to  produce the (quantitative) dynamics of collapse in
the realistic
(experimental) situation of Ref. \cite{exp5}. Instead, in this paper we
consider a
spherically-symmetric model to see if this model can describe the
essentials of the observed collapse
dynamics \cite{exp5}. We also present
results on central probability density of the condensate to confirm the
collapse. 

As we shall not be interested in a particular
boson-fermion 
system in this
paper, but will be concerned with the collapse of the fermionic condensate
in general we take in the rest of this paper $m_B=3 m_F=
m({\mbox{Rb}})$, whence $m_R=3m_F/4$. In the above two experimental
situations \cite{exp4,exp5} 
$m_B\approx 
3m_F$. 
Also,  we take $V_B({\bf r})=V_F({\bf r})= m_B\omega_B^2
r^2/2$ which corresponds to a reduction of $\omega_B$ by a factor
$\sqrt{m_B/m_F}$ as in the study by Modugno {\it et al.} \cite{zzz}. These
two
assumptions give a simpler analytical form of  the final equations we
derive
eliminating factors
of masses and frequencies without any consequence to our qualitative
study. 
Now  transforming to
dimensionless variables
defined by $x =\sqrt 2 r/l$,     $\tau=t \omega, $
$l\equiv \sqrt {\hbar/(m_B\omega_B)}$,
and
\begin{equation}\label{wf}
\frac{ \varphi_i(x;\tau)}{x} =  
\sqrt{\frac{4 \pi l^3}{N_i\sqrt 8}}\psi_i(r;t),
\end{equation}
$i=B,F$, 
we get
\begin{eqnarray}\label{d1}
&\biggr[&-i\frac{\partial
}{\partial \tau} -\frac{\partial^2}{\partial
x^2} 
+\frac{x^2}{4} 
+ 2 \sqrt 2    {\cal N}_{BB}\left
|\frac {\varphi_B}{x}\right|^2\nonumber \\          
&+& 4 \sqrt 2    {\cal N}_{BF}\left
|\frac {\varphi_F}{x}\right|^2 - i\xi_B  {\cal N}_{BB}  {\cal
N}_{BF} \left|\frac
{\varphi_B}{x}\right|^2  \left|\frac
{\varphi_F}{x}\right|^2
 \biggr]\nonumber \\  &\times & \varphi_B({ x};\tau)=0,
\end{eqnarray}
\begin{eqnarray}\label{d2}
&\biggr[&-i\frac{\partial
}{\partial \tau} -\frac{\partial^2}{\partial
x^2} 
+\frac{x^2}{4} 
+3\left(\frac{3\pi N_F}{2} \right)^{2/3}
\left|\frac {\varphi_F}{x}\right|^{4/3}\nonumber \\   
&+& 4 \sqrt 2    {\cal N}_{FB}\left
|\frac {\varphi_B}{x}\right|^2    
- i\xi_F  {\cal N}_{BB}^2\left|\frac {\varphi_B}{x}\right|^4
 \biggr]\nonumber \\ &\times& \varphi_F({ x};\tau)=0,
\end{eqnarray}
where
$ {\cal N}_{BB} =   N_B a_{BB} /l,$
$ {\cal N}_{BF} =   N_F a_{BF} /l,$ 
$ {\cal N}_{FB} =   N_B a_{BF} /l,$
$\xi_B=K_3/(2\pi^2 a_{BB}a_{BF}l^4\omega_B),$ and
$\xi_F=K_3/(2\pi^2 a_{BB}^2 l^4\omega_B)$. In the nonabsorptive case
$\xi_F=\xi_B=0$, the normalization of the wave-function components is
given by $\int_0^\infty dx |\varphi_i(x;\tau)|^2 =1, i=B,F.$ In the
absorptive case $\xi_F\ne 0$ and $\xi_B \ne 0$, the normalization 
reduces with time due to loss of atoms.

\section{Numerical Result}

We solve the coupled mean-field-hydrodynamic  equations (\ref{d1}) and
(\ref{d2}) numerically using a time-iteration
method based on the Crank-Nicholson discretization scheme
elaborated in Ref. \cite{sk1}.  
We
discretize the GP equation
using time step $0.00025$ and space step $0.05$ 
 spanning $x$ from 0 to 25. This
domain of space was sufficient to encompass  the whole condensate wave
function during collapse and expansion.

First we solve Eqs. (\ref{d1}) and (\ref{d2}) with
$\xi_B=\xi_F=0$. 
This
will allow us to find a stable bound state of
boson-fermion mixture.  After some experimentation
we take in our calculation $a_{BB} = 5$ nm,
$a_{BF} = -12.5$ nm, $\omega_B= 2\pi \times 100$ Hz, $N_B=4800$, and
$N_F=1200,$ so that $l\approx 1$ $\mu$m, 
$a_{BB}/l= 0.005$, $a_{BF}/l=-
0.0125$, $ {\cal N}_{BB} = 24,$ $ {\cal N}_{BF} = -15,$ and $ {\cal
N}_{FB} = -60.$ These values of parameters are similar to those employed
in experiments.
The dimensionless unit of time corresponds to
$\omega_B^{-1}\approx 1.6$ ms, and dimensionless unit of length
corresponds to $l/\sqrt 2 \approx 0.7$ $\mu$m.

\begin{figure}
 
\begin{center}
\includegraphics[width=0.9\linewidth]{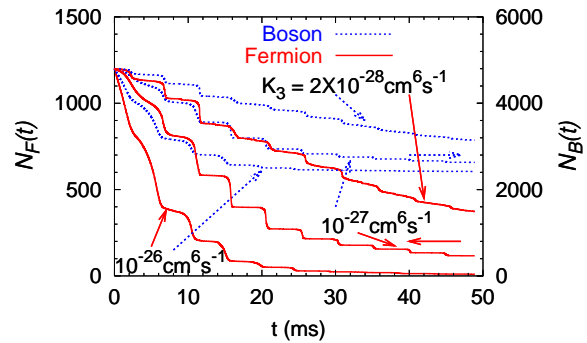}
\end{center}

\caption{(Color online) The evolution of the numbers of bosons $N_B(t)$
and fermions
$N_F(t)$ during collapse initiated by a jump in boson-fermion scattering
length $a_{BF}$ from $-12.5$ nm to $-37.5$ nm  for 
$K_3= 2\times 10^{-28}$ cm$^6$s$^{-1}$, $K_3= 10^{-27}$
cm$^6$s$^{-1}$, and 
$K_3=  10^{-26}$ cm$^6$s$^{-1}.$ The three upper (blue, dotted) curves
refer to
number of bosons and three lower (red, solid) curves refer to number of
fermions. The
curves are labeled by their respective $K_3$ values.  }

\end{figure}

Now we consider the collapse of fermions initiated by a sudden jump in the
boson-fermion scattering length from $a_{BF}=-12.5$ nm to $-37.5$ nm which
can be implimented near a boson-fermion Feshbach resonance. Such
boson-fermion Feshbach resonances between $^{23}$Na (boson) and $^6$Li
(fermion) atoms \cite{ket} and between $^{87}$Rb (boson) and $^{40}$K
(fermion) atoms \cite{jin} have been experimentally observed.  These
resonances should enable experimental control of the interspecies
interactions \cite{jin} and hence can be used to increase the attractive
force between bosons and fermions which in turn increases the attractive
nonlinearities $4\sqrt 2 {\cal N}_{BF}$ and $4\sqrt 2 {\cal N}_{FB}$ in
Eqs. (\ref{d1})  and (\ref{d2}). If these attractive nonlinear terms are
larger than the repulsive nonlinearities in these equations it is possible
to have collapse of fermions or bosons or both. In order to have collapse,
the effective nonlinearities in these equations should be attractive.  

Due to the imaginary terms in Eqs. (\ref{d1}) and (\ref{d2}) the numbers 
of bosons and fermions decay with time. When the net
nonlinear attraction
in these equations is small there is smooth and steady decay of number of
atoms. However, when the  net
nonlinear attraction is gradually increased the steady decay of number of
atoms develops into a violent decay called collapse. When this happens the
condensate loses a significant fraction of atoms in a small interval of
time (milliseconds) after which a remnant condensate with a reasonably
constant number of atoms is formed. Also, during and immediately
after collapse, the wave function of the condensate becomes very unsmooth
and spiky in nature as opposed to a reasonably smooth wave function in the
case of a steady decay.

We study the the evolution of the boson and fermion numbers in the
condensate from time $t=0$ to $t=50$ ms after the sudden jump in the
scattering length from $a_{BF}=-12.5$ nm to $-37.5$ nm at time $t=0$. The
evolution of boson and fermion numbers depends on the value of the loss
rate $K_3$. We study the sensitivity of the result on $K_3$ by
performing the calculation for different loss rates. In Fig. 1 we plot the
evolution of the boson and fermion numbers for different loss rates: $K_3
=2\times 10^{-28}$ cm$^6$s$^{-1}$, $10^{-27}$ cm$^6$s$^{-1}$, and
$10^{-26}$ cm$^6$s$^{-1}$. 
With the increase of $K_3$ the rate of decay
increases, although the results for different $K_3$ are qualitatively
similar. We see in Fig. 1 that in all cases 
both the number of bosons and fermions decay
rapidly and attain an approximately  constant (remnant)  number in less
than 50
ms. The
panorama is similar to the collapse in attractive bosonic condensate
studied experimentally by Donley {\it et al.} \cite{don}, where also a
cold remnant bosonic condensate is formed at the end of the collapse. 
In the recombination process (\ref{g}) two bosonic and one fermionic atoms
are lost. In Fig. 1  we find that during the collapse about 2000
bosonic atoms and 1000 fermionic atoms are lost.

\begin{figure}
 
\begin{center}
\includegraphics[width=0.7\linewidth]{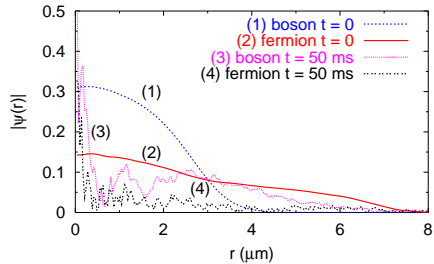}
\end{center}

\caption{(Color online) The initial ($t=0$) and
final ($t=50$ ms)  fermion and boson wave functions in arbitrary units
before and after collapse   initiated by a jump in boson-fermion
scattering
length $a_{BF}$ from $-12.5$ nm to $-37.5$ nm. }

\end{figure}

Experimentally, Modugno {\it et al.}
measured the following loss rate $K_3= 2 (1) \times 10 ^{-27}$
cm$^6$s$^{-1}$
for the $^{40}$K-$^{87}$Rb system \cite{exp5}.
In the remainder of this paper   we shall use the value $K_3
= 10^{-27}$ cm$^6$s$^{-1}$ for all densities and all  values of
scattering
lengths. This value of $K_3$ gives a loss rate during collapse compatible
with experiment \cite{exp5}, where a significant fraction of fermions are
lost in an interval of time $\Delta <50$ ms.
We further examined the
wave function of the bosons and fermions  to determine whether the
decay  in Fig. 1  really corresponds to a collapse and not to a slow
evaporation  due to the imaginary terms in Eqs. (\ref{d1}) and
(\ref{d2}).

\begin{figure}
 
\begin{center}
\includegraphics[width=0.7\linewidth]{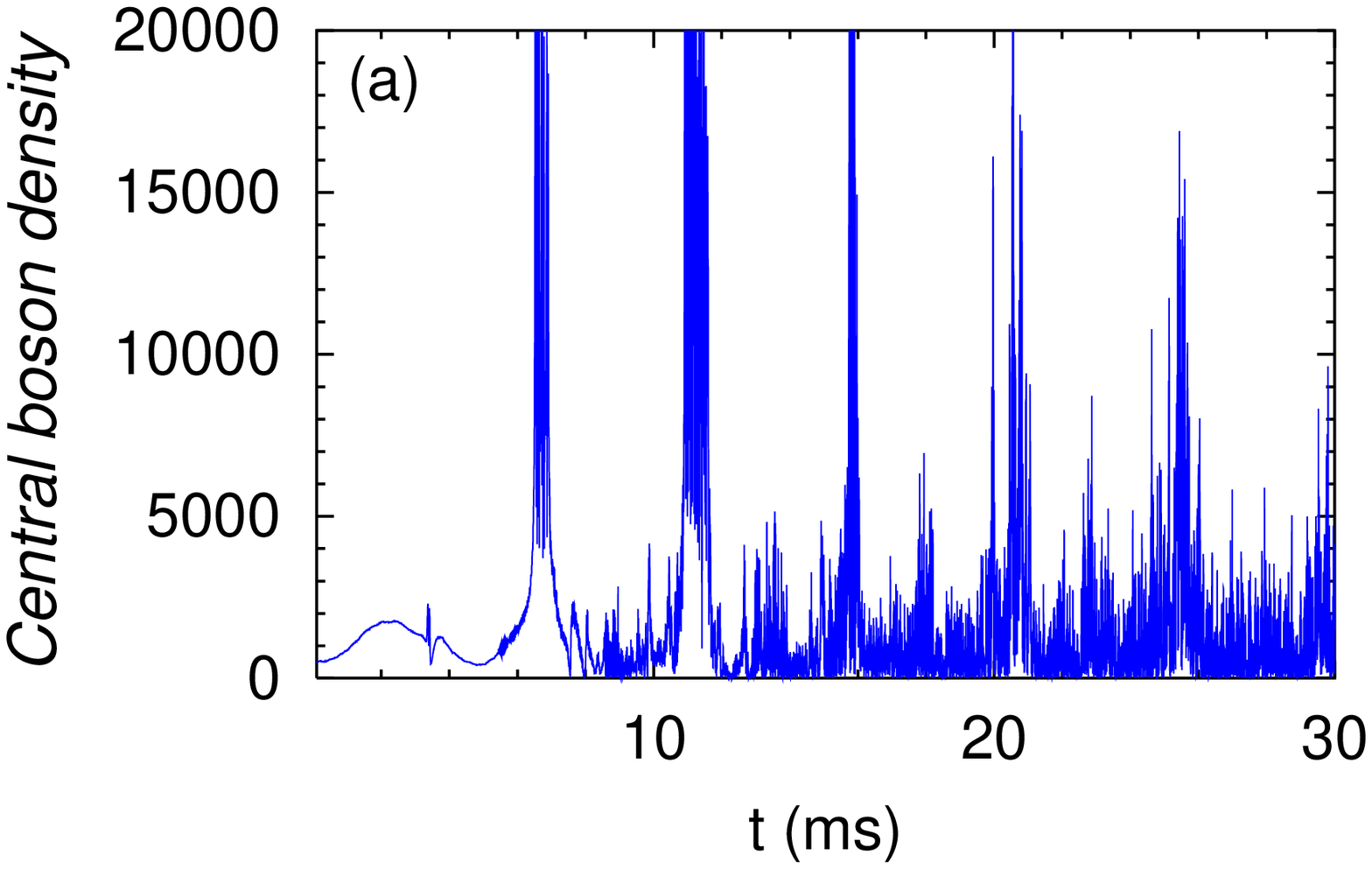}
\includegraphics[width=0.7\linewidth]{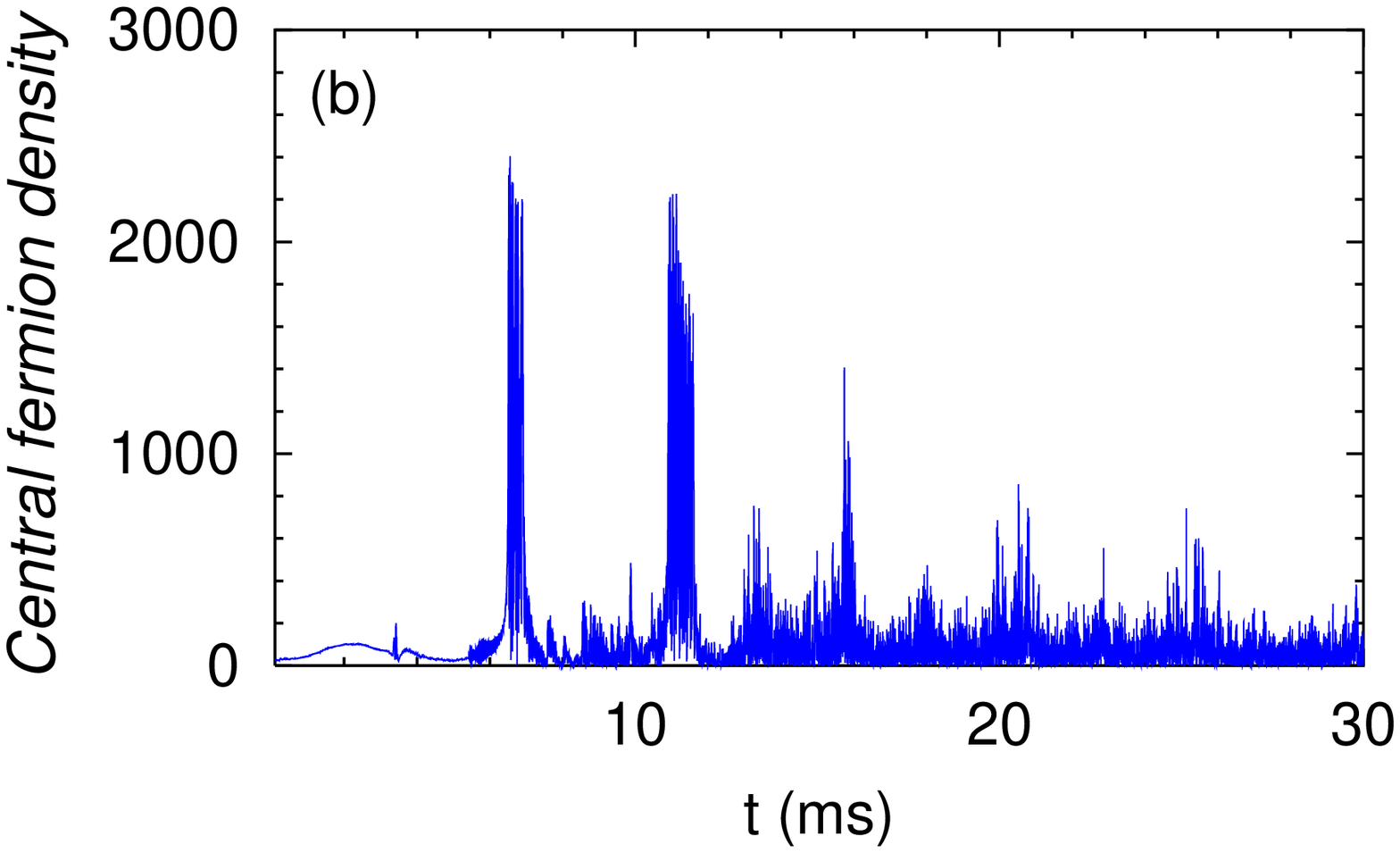}
\end{center}

\caption{Evolution of the (a)  central boson probability density
($N_B|\psi_B(0,t)|^2$) and (b)  central fermion probability density 
($N_F|\psi_F(0,t)|^2$) in arbitrary units during the collapse exhibited in
Figs. 1 and 2.}

\end{figure}

 In Fig. 2
 we plot the profiles of the bosonic and fermionic wave functions in
arbitrary units
at times $t=0$ and $t=50$ ms.
A close look at Fig. 2 reveals that before collapse at $t=0$ the
bosonic and fermionic wave functions are smooth and the fermionic wave
function extends over a larger distance than the bosonic wave
function. This is due to the large repulsion between the fermions in the
spin polarized state due to the Pauli exclusion principle. Hence the
bosonic condensate lies well inside the fermionic condensate. This was
noted in experiment \cite{exp5}, as well as in previous theoretical
studies \cite{capu,yyy,zzz}.
The wave functions after collapse  have an entirely different profile. 
As expected the wave functions are highly peaked in the central ($r=0$) 
region and develop spikes. 
However, they extend over a large distance
too. The final spiky wave function indicates the collapse in contrast to a
smooth final  wave function corresponding to a  steady loss of
atoms. The collapse is a quick process lasting at most a few tens of
milliseconds when a significant fraction of atoms are lost. For example,
in Fig. 1 for $K_3=10^{-27}$ cm$^6$s$^{-1}$, the collapse lasts for the
first 
25 ms when most of the atoms are lost. After this 
interval the rate of loss of atoms is reduced and 
remnant bosonic and fermionic condensates 
with a roughly constant number of atoms are formed. 
 The wave
function after the collapse of a purely bosonic condensate also exhibits a
similar behavior \cite{th2}. 

\begin{figure}
 
\begin{center}
\includegraphics[width=0.9\linewidth]{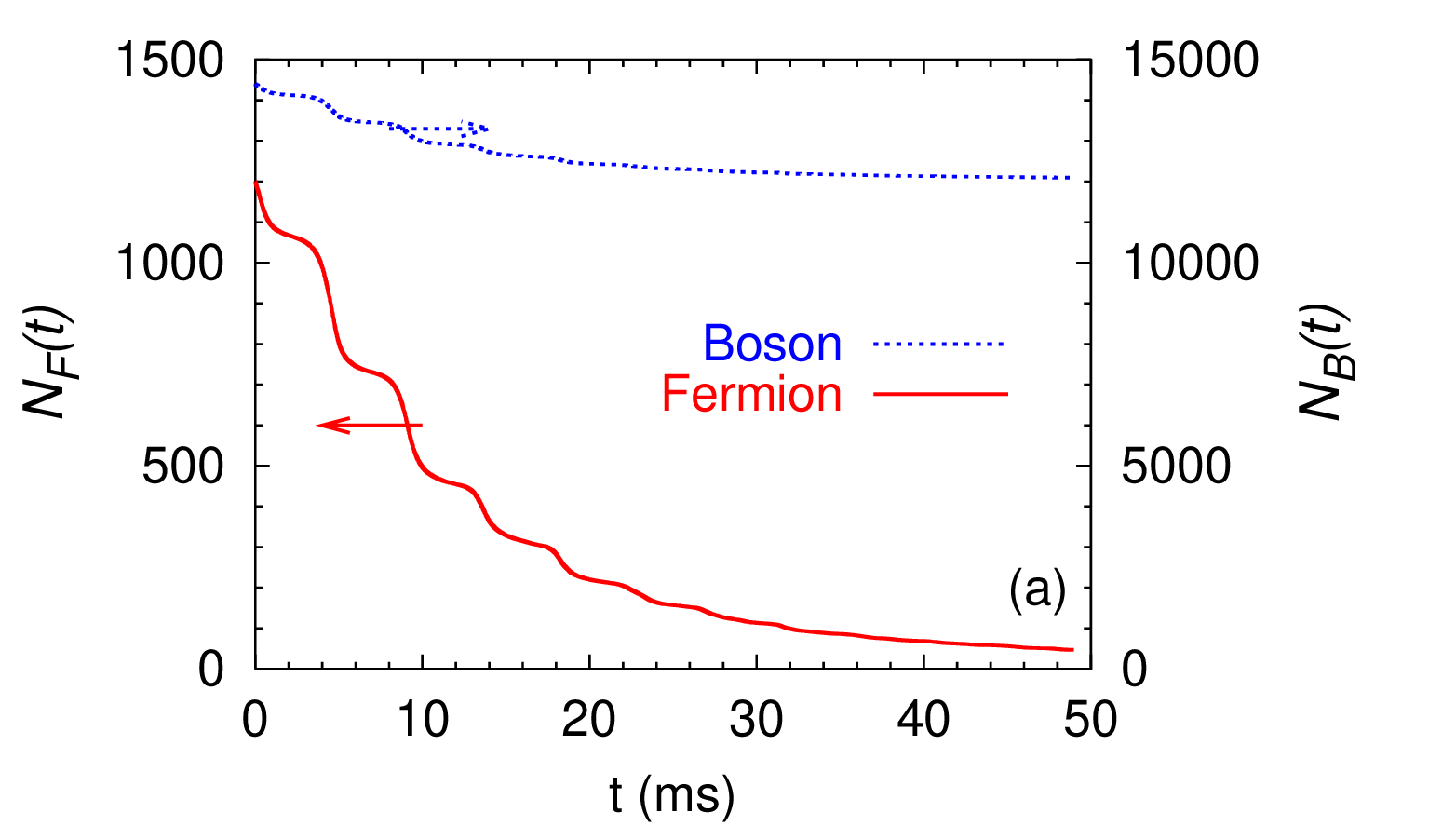}
\includegraphics[width=0.9\linewidth]{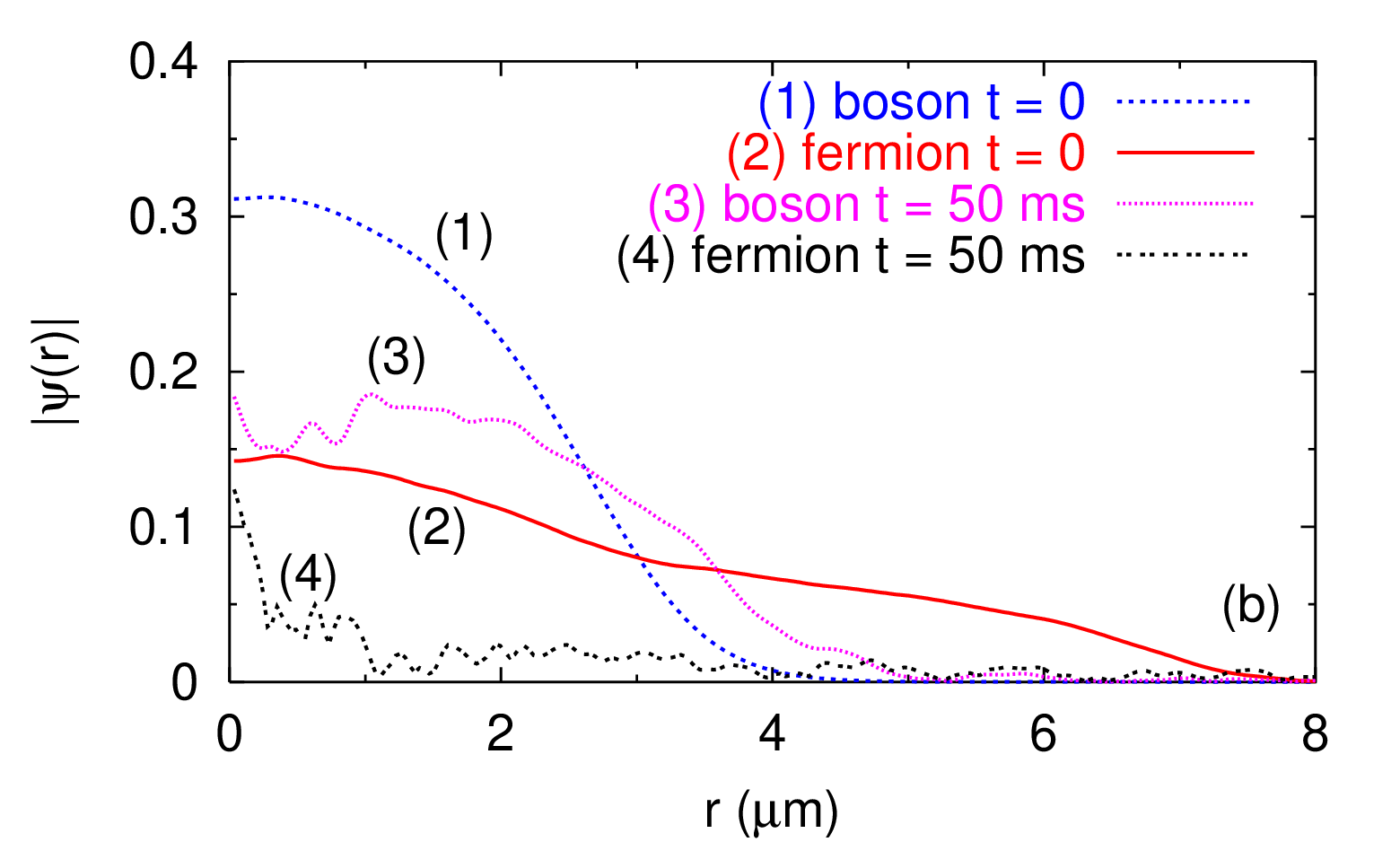}
\end{center}

\caption{(Color online) (a) The evolution of the numbers of bosons
$N_B(t)$ and fermions
$N_F(t)$ during collapse initiated by a jump in the boson number from 4800
to 14400 for $K_3= 10^{-27}$ cm$^6$s$^{-1}$. (b) The initial ($t=0$) and
final ($t=50$ ms)  fermion and boson wave functions in arbitrary units 
before and after the collapse in (a). }

\end{figure}

To confirm further the  collapse in Figs. 1 and 2 for $K_3=
10^{-27}$
cm$^6$s$^{-1}$, we plot 
in Fig. 3  the evolution of the central probability  density of
(a) the bosonic
($N_B|\psi_B(0,t)|^2$)
and (b) fermionic ($N_F|\psi_F(0,t)|^2$)  
components in arbitrary units during collapse. 
We note a very strong fluctuation of the central density reminiscent of
collapse both in the bosonic and fermionic components. Similar variations
are common to the collapse of a purely bosonic condensate \cite{th2}. Such
strong fluctuation of the central density  could not be due to a weak
evaporation of the condensate due to recombination (\ref{g}).

Next we consider fermionic collapse initiated by an increase of the 
number of bosonic
atoms. 
In Fig. 4 (a) we plot the evolution of the boson and fermion numbers in 
the condensate from time $t=0$ to $t=50$ ms
after the  jump in the  boson number  from
4800 to 14400 at time $t=0$. In this case there is a decay in boson and
fermion numbers. The decay of boson numbers ($\sim 2000$) is double the
decay of fermion numbers ($\sim 1000$). To determine if this case
corresponds to collapse,  in Fig. 4
(b) we plot the profiles of the bosonic and fermionic wave functions in
arbitrary units
at times $t=0$ and $t=50$ ms. At $t=50$ ms the fermionic wave function is
sharply
peaked at the center and spiky in nature, whereas the reasonable
smooth bosonic wave function does not exhibit
any central peaking. 
The slowly varying final bosonic component
corresponds to a weak loss of atoms and not to collapse. The collapse of
the
fermionic component deserves further examination. 
In Fig. 5 we plot the evolution of the central probability  density of
the  fermionic ($N_F|\psi_B(0,t)|^2$)  
component in arbitrary units during collapse exhibited in Figs. 4. 
The central density exhibits rapid oscillation indicating
collapse. However, the fluctuation in Fig. 5 is less than those noted in
Figs. 3 indicating a less violent collapse in this case. This is quite
reasonable as in Fig. 2 both components undergo collapse.

The initial nonlinearities in Eqs. (\ref{d1}) and (\ref{d2}) are ${\cal
N}_{BB}=24$, ${\cal N}_{BF}=-15$, ${\cal N}_{FB}=-60$, whereas the final
nonlinearities in Figs. 1 are ${\cal N}_{BB}=24$, ${\cal N}_{BF}=-45$,
${\cal
N}_{FB}=-180$. The final attractive (negative) nonlinearities are so
strong that the effective nonlinearities in both Eqs. (\ref{d1}) and
(\ref{d2}) have become attractive and large in nature. This is responsible
for the collapse observed in both components in Figs. 1 and 2. In the
situation of Fig. 4 the final
nonlinearities are  ${\cal N}_{BB}=72$, ${\cal N}_{BF}=-15$,
${\cal
N}_{FB}=-180$. We find that, as a result, the repulsive nonlinearity in
the
bosonic equation 
 (\ref{d1}) has increased and hence the bosonic condensate does not
collapse; whereas the attractive nonlinearity in the fermionic equation 
(\ref{d2}) has increased by a factor of 3 and hence this component
undergoes collapse. 
However, in Fig. 4 (a)  there is a steady loss of atoms in the repulsive  
bosonic condensate during the collapse of the fermionic condensate.

\begin{figure}
 
\begin{center}
\includegraphics[width=0.9\linewidth]{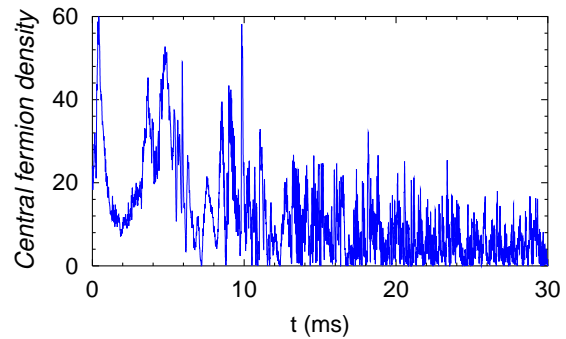}
\end{center}

\caption{Evolution of the  central fermion probability density 
($N_F|\psi_F(0,t)|^2$) in arbitrary units during the collapse exhibited in
Figs. 4.}

\end{figure}

The present study indicates that under different situations the fermionic
component in a trapped condensed boson-fermion mixture can undergo a
collapse due to an attractive  boson-fermion interaction. In the present
study we consider two  situations. Of these, the
fermionic collapse associated with the increase in the number of bosons
has been exploited in the experiment of Modugno {\it et al.} \cite{exp5}.
In the present model
simulation, we
introduced a sudden jump in the number of atoms which is not possible in
the laboratory. In the experiment of Ref. \cite{exp5} a steady increase in
boson numbers has been used to initiate the fermionic collapse.
However, the situation of collapse exploiting a Feshbach resonance 
in the boson-fermion system is more exciting and controlled experiment
could
be done by jumping the boson-fermion scattering length  $a_{BF}$     
as in  the
collapse experiment with bosonic condensate \cite{don}.  This seems
possible in $^6$Li-$^{23}$Na \cite{ket} and $^{40}$K-$^{87}$Rb \cite{jin} 
systems using the
recently discovered Feshbach
resonances in these systems. The increase in
the number of bosonic atoms in the boson-fermion condensate to initiate
the collapse is a slow stochastic process as opposed to sudden
controlled jump in the
scattering length leading to a much violent collapse. In this connection
we recall that the collapse of a purely bosonic  $^{7}$Li condensate in
Ref. \cite{hulet} by a steady increase of the atom number 
was a  stochastic process, whereas the collapse of $^{85}$Rb
atoms in Ref. \cite{don} by a jump in scattering length was a more
violent,
nevertheless more exciting, process.  
However, the
collapse initiated by a jump in the  scattering length $a_{BF}$
will
require a careful study of the boson-fermion system in the search of an
appropriate Feshbach resonance \cite{ket,jin}.

An examination of Eqs. (\ref{d1}) and (\ref{d2}) with $\xi_F=\xi_B=0$
reveals that for a fixed $N_F$ 
with the increase of the boson number $N_B$ the attractive nonlinearity 
${\cal N}_{FB}$ can overcome the repulsive Fermi pressure 
and lead to a
collapse of the fermionic condensate. Hence, for a
fixed  $N_B$ and fixed values of scattering lengths, $N_F$ has to be
larger than a  critical value in order to have a stable condensate. The
fermionic
collapse is a function of the boson-fermion scattering length $a_{BF}$ and 
the boson number $N_B$. These aspects of collapse have been studied via
equilibrium equations (\ref{c}) and (\ref{d}) by Modugno {\it et al.}
\cite{zzz}.      Finally, we comment that a large number of fermions $N_F$
in  Eqs. (\ref{d1}) and (\ref{d2}) will lead to a strong attraction in the
bosonic equation  (\ref{d1}) via the term ${\cal N}_{BF}$ which may
initiate a collapse of the bosonic atoms. An increase of  $N_B$
stabilizes the bosonic condensate but may initiate collapse in  the
fermionic one,
whereas
an increase in $N_F$ stabilizes the fermionic condensate
but may start
collapse
in the
bosonic one.  Finally, an increase in $|a_{BF}|$ may initiate collapse in
both components. 
A detailed study of these
features in the actual experimental situation would be a welcome future
work.

\section{Summary}
 
We have suggested a coupled set of time-dependent mean-field-hydrodynamic
equations for a trapped boson-fermion condensate using the
(time-independent) energy functional of Ref. \cite{capu} successfully used
to study the equilibrium states of the trapped boson-fermion condensate.
One could identify collapse from the time independent formulation, where
equilibrium stationary solutions for the boson-fermion mixture cease to
exist.  The present time-dependent generalization permits us to study
nonequilibrium dynamics of the coupled boson-fermion condensate.
We use the time-dependent nonlinear model to study the collapse dynamics
for the boson-fermion mixture. There are two possibilities for the
collapse of the fermionic component. From Eq. (\ref{d2}) we see that this
happens when the attractive nonlinearity ${\cal N}_{FB}$ becomes stronger
either via an increase in boson number $N_B$ or via an increase in the
strength of boson-fermion interaction $|a_{BF}|$. We considered both
possibilities in the present numerical study. The increase of  $|a_{BF}|$
leads to a collapse of both components, whereas the increase of  $N_B$
leads to a collapse of the fermionic component alone. The collapse is more
violent in the first case. The present simulation was done in a
spherically symmetric model. The collapse dynamics is strongly dependent
on the loss rate $K_3$. Once one has a good knowledge of the variation of
$K_3$ with    $|a_{BF}|$, and high-quality experiment data will
be available, a more realistic calculation will be worth doing in the
future.

 
\acknowledgments

The work is supported in part by the CNPq 
of Brazil.


\end{document}